# Equivalent electric circuit of a carbon nanotube based molecular conductor

ChiYung Yam, Yan Mo, Fan Wang, Xiaobo Li, GuanHua Chen[1]
Department of Chemistry
Centre of Theoretical and Computational Physics
University of Hong Kong, Hong Kong

Xiao Zheng[2]
Department of Chemistry, The University of Hong Kong
Department of Chemistry, Hong Kong University of Science and Technology
Hong Kong

Yuki Matsuda, Jamil Tahir-Kheli, and William A. Goddard III[3]
Materials and Process Simulation Center (MC 139-74)
California Institute of Technology, Pasadena CA, 91125

Abstract

We apply our first-principles method to simulate the transient electrical response through carbon nanotube based conductors under time-dependent bias voltages, and report the dynamic conductance for a specific system. We find that the electrical response of the carbon nanotube device can be mapped onto an equivalent classical electric circuit. This is confirmed by studying the electric response of a simple model system and its equivalent circuit.

As the Moore's Law Roadmap for semiconductor industry is followed into the scale of 20 nm, it becomes important to understand dynamic response of nanometer scale molecular electronic devices.[1-14] This requires the use of quantum mechanics to ensure the proper treatment of transient and quantum effects. For practical use by design engineers, it is crucial to cast these quantum effects into the form of classical electric circuits. An important question is whether such a mapping is possible, and if so, what are the forms of the equivalent circuits?

As potentially important components of next-generation integrated circuits, carbon nanotubes (CNTs) have been studied extensively.[3-9,15] High frequency electrical response of micrometer-long individual and bundled CNTs have been measured,[5] and equivalent electric circuits have been proposed.[5,8-9] In this work, we concentrate on a nanometer-scale CNT-based electronic device, and apply first-principles quantum mechanics (QM) to determine its dynamic electrical response. Our system is a (5,5) CNT (0.68 nm in diameter and 0.62 nm in length) which is bonded covalently between two aluminum electrodes and shown in Fig. 1a.

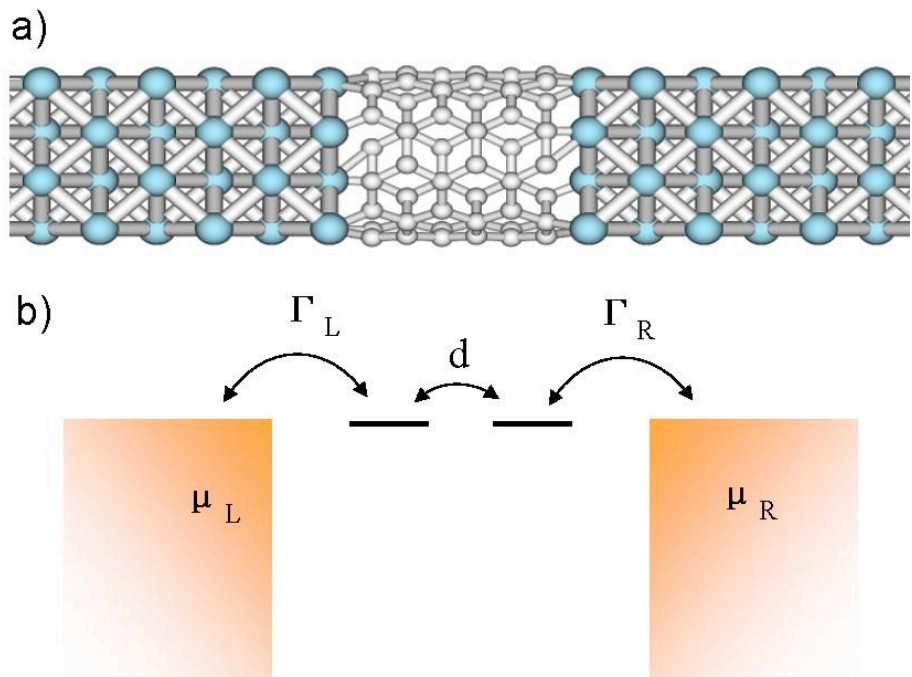

Fig.1 (a) Prototype used for explicit QM calculations of a carbon nanotube based conductor: the (5,5) CNT device with aluminum electrodes. (b) A two-site system coupled to the left and right electrodes. $\Gamma_L = \Gamma_R = \gamma/2$

To predict the transient electrical response of this molecular device, we use the rigorous time-dependent density functional theory (TDDFT) that we developed recently[16] to evaluate the time-dependent current through the device. We included explicitly in the simulation box 48 Al atoms of each electrode along with 60 C atoms of the CNT.

Figures 2a and 2b show the current versus time for two different types of bias voltage switched on at t = 0. In Fig. 2a bias voltage $V_b$ is turned on exponentially. We observe that the current reaches its steady state in 12 fs. The time dependent current can be fitted by $I_0(1-e^{-t/\tau})$ with $\tau$ = 2.8 fs and $I_0$ =13.9 nA, leading to a characteristic time of 2.8 fs. The reason for such a fast switch-on time is that the process involves only electrons. Figure 3a plots the potential energy change for an electron along the central axis at t = 0.02, 1 and 12 fs. The potential change is the sum of the applied potential and the potential caused by the induced charge. Our calculation leads to following observations:

- After turning on the bias voltage, at t = 0.02 fs the electrons have not yet responded to the applied voltage, and the external field is hardly screened, dropping uniformly across entire Al-CNT-Al system.
- At t = 1 fs the potential drop occurs mostly on the CNT since aluminum is more polarizable. It takes less than 1 fs for the electrons on the Al electrodes to screen the applied potential.
- In Fig. 3b we plot induced charge along Al-CNT-Al at t = 4 fs. The red indicates positive charge while the blue indicates the negative charge. Alternating positive and negative charge distributions on CNT cancel each other so that its net induced charge is zero. The excess charge resides primarily at two interfaces and forms an effective capacitor as depicted schematically in Fig. 3c.

We also considered the response to a sinusoidal bias voltage turned on at t = 0. Fig. 2b shows the corresponding time-dependent current. We see a phase delay in the current response to bias voltage. This implies at this frequency the device is overall inductive.

The Al-CNT-Al system is symmetric. As a consequence, there is no net charging of the device, and the time dependent current is conserved.[17] This was confirmed by our numerical simulation. The current entering the system has the same magnitude as the current leaving as shown in Figs. 2a and 2b. Therefore, the conductance matrix element $G_{\alpha\beta}(\omega)$ ($\alpha, \beta$ = L or R) satisfies $G_{LL} = G_{RR} = -G_{LR} = -G_{RL} = G(\omega)$.[17,18] Taking the Fourier transform, the simulated transient dynamics leads to $I(\omega)$ and $V(\omega)$, from which we obtain the dynamic conductance $G(\omega) = I(\omega) / V(\omega)$. We find that both types of bias voltages lead to essentially the same dynamic conductance. This implies that the electrical response is in linear response regime and also validates the accuracy of our calculations. Figure 2c shows the real and imaginary parts of the resulting dynamic conductance.

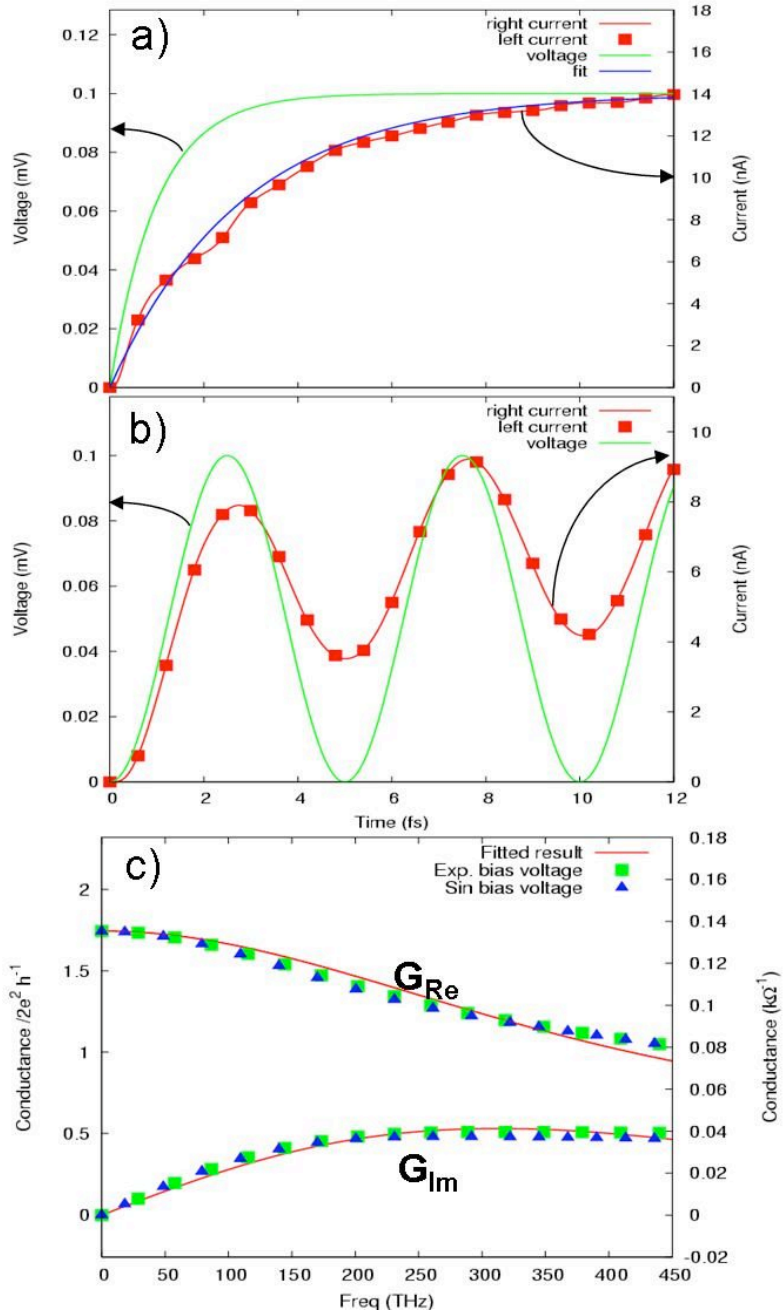

Fig. 2 (a) and (b) Transient current (red lines and squares) and applied bias voltage (green lines) for Al-CNT-Al system. (a) Bias voltage is turned on exponentially, $V_b = V_0(1-e^{-t/a})$ with $V_0 = 0.1$ mV and a time constant $a = 1$ fs. Blue line in (a) is a fit to transient current. (b) Bias voltage is sinusoidal with a period of 5 fs. Red line is for current from right electrode, and squares are current from left electrode. (c) Dynamic conductance calculated from the exponential bias voltage turned on at t=0 (solid squares) and the sinusoidal bias voltage (solid triangle) turned on at t=0. The red lines are the fitted results. The upper curves are the real part of the conductance while the lower ones are the imaginary part.

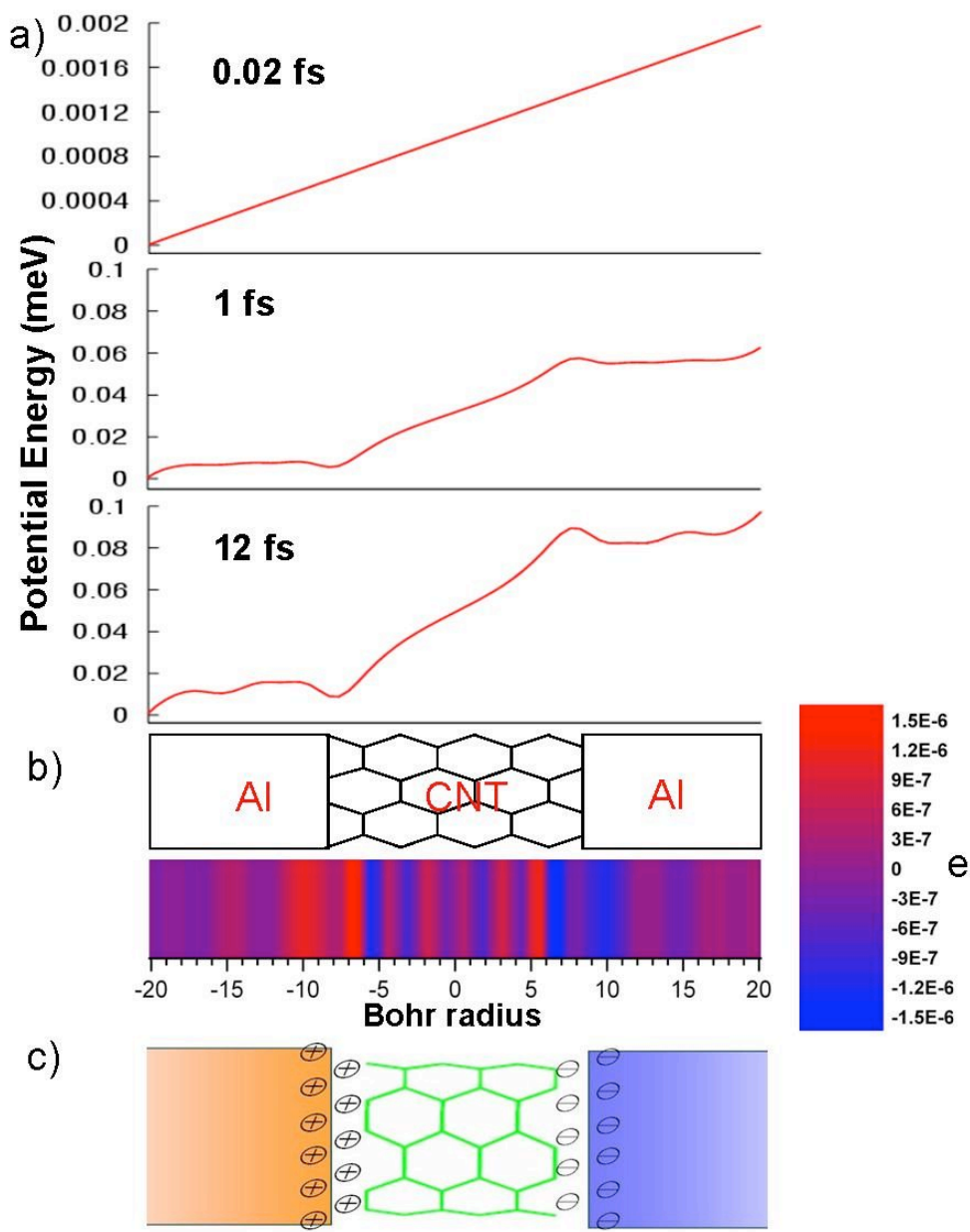

Fig. 3. (a) Electrostatic potential energy distribution along the central axis at t = 0.02, 1 and 12 fs. (b) Charge distribution along Al-CNT-Al at t = 4 fs. (c) Schematic diagram showing the induced charge accumulation at two interfaces, which forms an effective capacitor.

Now the question is how to model the Al-CNT-Al device. Our above results and analysis show that our molecular device has both inductive and capacitive components. As the current enters into the device region from the left electrode, a part of it, $I_c$, charges the left interface (see Fig. 4a). The remaining current, $I_L$, goes straight through the device and is joined by $I_c$ at the right interface. Therefore, our device can be modeled by the classical circuit depicted in Fig. 4b. At zero frequency, the steady current goes only through the R-L branch. $R_L$ is simply the steady state resistance [19,20], which we calculate to be 7.39 kΩ. A similar circuit was proposed to fit the numerical dynamic conductance of a model quantum wire [21] with an additional inductor to account for a resonance at a high frequency.

Büttiker and coworkers [10] studied a mesoscopic capacitor made of two plates with each coupled to an electron reservoir via a narrow lead. They discovered that the charge relaxation resistance $R_c$ is universal independent of transmission details, leading to $R_C = R_1 + R_2$ with $R_1$ and $R_2$ given by

$$R_\alpha = \frac{h}{2e^2} \frac{\sum_n D_{\alpha,n}(E_F)^2}{\left(\sum_n D_{\alpha,n}(E_F)\right)^2} \tag{1}$$

where $D_{\alpha,n}(E_F)$ is the density of states (DOS) at Fermi energy for $n$th spin-specific charging channel of plate α (α=1,2), and $\sum_n$ is over all charging channels for α. This was confirmed by a recent experiment. [22]

In our Al-CNT-Al system, two interfaces correspond to two plates of the capacitor (see Fig. 3c) and couple to the electrodes via Al leads. Our CNT has two degenerate orbitals for transmission. Both spin-up and spin-down electrons contribute to the transmission. Therefore, there are four charging channels for each interface, and according to Eq. (2), the charge relaxation resistance for the Al-CNT-Al system is

$$R_C = \frac{h}{2e^2} \times \frac{1}{4} + \frac{h}{2e^2} \times \frac{1}{4} = \frac{h}{4e^2}. \tag{2}$$

We tune the values of $L$ and $C$ to fit the calculated dynamic conductance while fixing $R_C = \frac{h}{4e^2}$ and $R_L$ = 7.39 kΩ. The resulting values of $L$ and $C$ are 16.6 pH and 0.073 aF, respectively. Red lines in Fig. 2c are the real and imaginary values of the dynamic conductance of the equivalent electric circuit, which agree well with our calculated dynamic conductance up to 450 THz. We can estimate the capacitance $C$ directly from the excess charge $Q$ at the interfaces. For bias voltage $V_b$ = 0.1 mV, we find that at the steady state $Q$ is ~0.03 e. Therefore, we estimate the capacitance as

Q/V$_b$ = 0.05 aF,

which is of the same magnitude as the calculated $C$ = 0.073 aF. Since there is uncertainty to define the excess charge at the interfaces, the value (0.05 aF) is meant to be an estimation only. According to Eqs. (5) & (7) below, the inductance $L$ is $\sim \left(\frac{h}{\gamma}\right) R_L$.

Average line width $\langle\gamma/4\rangle$ at the Fermi energy is ~0.38 eV. Thus, $\left(\frac{h}{\gamma}\right)R_L \approx 18.8$ pH. This is close to the calculated $L = 16.6$ pH.

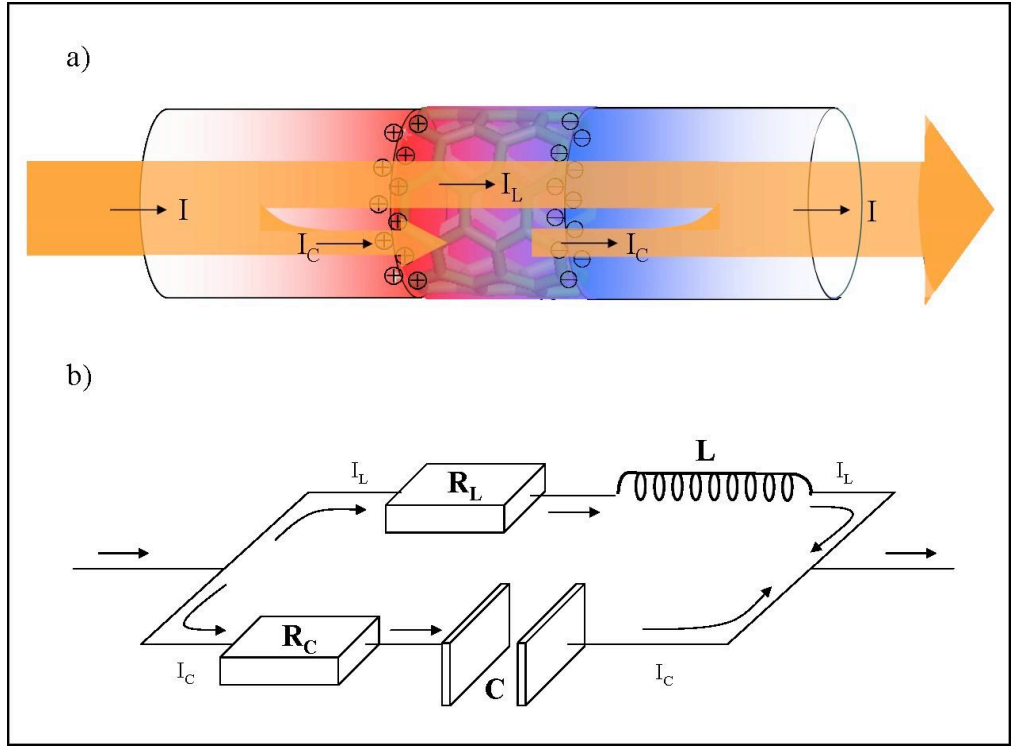

Fig. 4. (a) Current flow in the parallel circuit. (b) The equivalent electric circuit. $R_L = 7.39$ k$\Omega$, $L = 16.6$ pH, $R_C = 6.45$ k$\Omega$, and $C = 0.073$ aF.

To confirm further the equivalent electric circuit for our CNT based molecular conductor, we designed a simple model: a two-site system in contact with the left and right electrodes (see Fig. 1b). The two sites are degenerate in energy $\varepsilon_0$. For the capacitor limit where $d = \frac{\gamma}{4}\eta$, $0<\eta<<1$ and $\varepsilon_0=\mu_L=\mu_R$, we derive analytically the frequency dependent conductance as follows,

$$G(\omega) = \frac{2e^2}{h}\int dE \frac{f_\beta(E) - f_\beta(E+\hbar\omega)}{\hbar\omega} \mathrm{Tr}\left\{G^r_{eq}(E+\hbar\omega)\left[\hat{\Gamma}_\alpha - \frac{i}{2}\hbar\omega\cdot\hat{1}\right]G^a_{eq}(E)\hat{\Gamma}_\beta\right\}$$

$$= \frac{2e^2}{h}4\eta^2 - i\omega\frac{2e^2}{h}\frac{4\hbar}{\gamma}\left(1-8\eta^2\right) + \omega^2\frac{2e^2}{h}\left(\frac{4\hbar}{\gamma}\right)^2\left(1-\frac{32}{3}\eta^2\right) + O(\omega^3)$$

(3)

Expanding the dynamic conductance of the electric circuit in Fig. 4b in $\omega$ leads to a dynamic conductance:

$$G(\omega) = \frac{1}{R_L} + i\omega\left(-C + \frac{L}{R_L^2}\right) + \omega^2\left(R_C C^2 - \frac{L^2}{R_L^3}\right) + O(\omega^3) \tag{4}$$

Comparing Eqs. (3) and (4), we find that

$$\begin{aligned} R_L &= \frac{h}{8e^2}\left(\frac{\gamma}{4d}\right)^2, \\ L &= \left(\frac{8\hbar}{\gamma}\right) R_L, \\ C &= \frac{2e^2}{h}\frac{4\hbar}{\gamma}, \\ R_C &= \frac{h}{2e^2}. \end{aligned} \tag{5}$$

For the ballistic transport limit where $d = \frac{\gamma}{4}(1 \pm \delta)$ and $0 < \delta \ll 1$ and $\varepsilon_0 = \mu_L = \mu_R$, which is similar to our system, the dynamic conductance is

$$G(\omega) = \frac{2e^2}{h} + i\omega \frac{2e^2}{h}\frac{2\hbar}{\gamma}(1+\delta) - \omega^2 \frac{2e^2}{h}\left(\frac{2\hbar}{\gamma}\right)^2\left(1 + \frac{10}{3}\delta\right) + O(\omega^3) \tag{6}$$

Comparing Eq. (6) with Eq. (4), we find that

$$\begin{aligned} R_L &= \frac{h}{2e^2}, \\ L &= \left(\frac{2\hbar}{\gamma}\right) R_L\left(1 + \frac{5}{3}\delta\right) = \frac{2\hbar^2 \pi}{e^2 \gamma}\left(1 + \frac{5}{3}\delta\right), \\ C &= \frac{4}{3}\frac{\hbar}{\gamma} R_C^{-1}\delta = \frac{4}{3\pi}\frac{e^2}{\gamma}\delta, \\ R_C &= \frac{h}{2e^2}. \end{aligned} \tag{7}$$

The above calculation shows that near both capacitive and inductive limits the electrical response of two-site system can be modeled by the classical circuit in Fig. 4b, and its charging relaxation resistance $R_C$ is $\frac{h}{4e^2} + \frac{h}{4e^2} = \frac{h}{2e^2}$ which agrees with Eq. (1). If each site is two-fold degenerate, the charge relaxation resistance is reduced to half this value, *i.e.* $R_C = h/4e^2$, which is exactly the same as our Al-CNT-Al system.

An R-L circuit was proposed for long CNT with $L$ as the kinetic inductance.[5,8,9] In the presence of a substrate, extra parallel capacitors are introduced between the tube and substrate.[8,9] When a CNT sits on top of an electrode via van der Waals attraction, an effective capacitor is introduced between the CNT and electrode in addition to a parallel contact resistor.[5] The kinetic inductance and quantum capacitance of a long CNT are intrinsic properties of the tube, being determined by the DOS at the Fermi energy or the Fermi velocity $v_f$. In our case, the CNT is much shorter, and is welded to the electrodes covalently. The electrical responses of the interfaces and tube cannot be separated. Our inductance $L$ is determined by its self-energy due to the coupling to the electrodes or the dwell time of the conducting electron inside the device.[23] Our capacitance $C$ is mostly dictated by the local DOS at interfaces. As the length of CNT increases, the capacitance due to the interfaces decreases and the kinetic inductance of the tube dominates. As a result, our equivalent circuit is reduced to the R-L branch only, which is consistent with the equivalent electric circuit proposed for the long CNTs. [5,8,9] Wang and coworkers[23] introduced the dwell time $\tau_d$ to unify the inductance expression for short and long tubes as $L \sim \tau_d \frac{h}{e^2}$. For a long 1D system of length $l$, $\tau_d \sim l/v_f$, leading to an expression for the kinetic inductance per length $L/l \sim \frac{h}{v_f e^2}$.

Our nanoscale device has very small values of $L$ and $C$, leading to the short switching time. Such a fast switching speed for electronic devices based on nanometer-size CNTs indicates that these devices will not limit switching speeds in the foreseeable future. The equivalent electric circuit of the parallel R-L and R-C circuit in Fig. 4b is not limited to the CNT-based conductor studied in this work. It also applies to other two-terminal molecular, nanoscopic and mesoscopic electronic devices. $R_L$ is given by Landauer-Buttiker formula for steady state current,[19,20] and $R_c$ is the universal charge relaxation resistance depending only on the number of charging channels and spin polarization.[10] $L$ is determined by the dwell time of the electrons inside the device as $L \sim \tau_d \frac{h}{e^2}$.[23] $C$ is the electro-chemical capacitance which is determined by the geometry and the DOS at interfaces.[10,23] These results should be useful in designing the nanoscale electronics systems required over the next decade.

**Acknowledgments:** The authors thank Hong Guo, Jian Wang, and YiJing Yan for stimulating discussions. The Caltech team was supported partly by Intel Components Research (Portland OR) and by NSF (CCF-0524490). The HKU team was supported by the Hong Kong Research Grant Council (HKU 7011/06P, N_HKU 764/05, HKUST 2/04C).

[1]ghc@everest.hku.hk
[2]xzheng@yangtze.hku.hk
[3]wag@wag.caltech.edu

Reference:

1. A. Aviram and M.A. Ratner, Chem. Phys. Lett. **29**, 277 (1974).
2. S. Datta, *Electron Transport in Mesoscopic Systems* (Cambridge University Press, Cambridge, England, 1995).
3. J. Taylor, H. Guo and J. Wang, Phys. Rev. B **63**, 245407 (2001).
4. J. Xiang, W. Lu, Y. Hu, Y. Wu, H. Yan and C. M. Lieber, Nature **441**, 489(2006).
5. J. J. Plombon, K. P. O'Brien, F. Gstrein, V. M. Dubin and Y. Jiao, App. Phys. Lett **90,** 063106 (2007); L. Gomet-Rojas, S. Bhattacharyya, E. Mendoza, D. C. Cox, J. M. Rosolen, and S. R. P. Silva, NanoLett. **7**, 2672 (2007).
6. Y. H. Kim, J. Tahir-Kheli, P. A. Schultz and W. A. Goddard III, Phys. Rev. B **73,** 235419 (2006).
7. A. Javey, J. Guo, M. Paulsson, Q. Wang, D. Mann, M. Lundstrom and H. J. Dai, Phys Rev. Lett. **92,** 106804 (2004).
8. P. J. Burke, IEEE Trans. Nano. **2**, 55 (2003); S. Li, Z. Yu, S.-F. Yen, W. C. Tang, and P. J. Burke, NanoLett. **4**, 753 (2004).
9. A. Raham, J. Guo, S. Datta, and M. S. Lundstrom, IEEE Trans. Electron Devices **50**, 1853 (2003).
10. M. Büttiker, H. Thomas and A. Pretre, Phys. Lett. A **180**, 364 (1993); Y. M. Blanter, F. W. J. Hekking, and M. Büttiker, Phys. Rev. Lett. **81**, 1925 (1998).
11. K. Burke, R. Car, and R. Gebauer, Phys. Rev. Lett. **94**, 146803 (2005).
12. S.-H. Ke, H. U. Baranger and W. Yang, Phys. Rev. B **70**, 085410 (2004).
13. Y.-G. Yoon, P. Delaney and S. G. Louie, Phys. Rev. B **66,** 073407 (2002).
14. A. Di Carlo, M. Gheorghe, P. Lugli, M. Sternberg, G. Seifert, and T. Frauenheim, Physica B **314**, 86 (2002).
15. X. Zheng, G.H. Chen, Z.B. Li, S.Z. Deng, N.S. Xu, Phys. Rev. Lett. **92**, 106803 (2004).
16. X. Zheng, F. Wang, C.Y. Yam, Y. Mo and G. H. Chen, Phys. Rev. B **75** 195127 (2007).
17. M. Büttiker, A. Pretre and H. Thomas, Phys. Rev. Lett. **70**, 4114 (1993); Phys. Rev. Lett. **71**, 465 (1993); B. Wang, J. Wang and H. Guo, Phys. Rev. Lett. **82,** 398 (1999).
18. Y. Fu and S. C. Dudley, Phys. Rev. Lett. **70**, 65 (1993); Phys. Rev. Lett. **71**, 466 (1993).
19. R. Landauer, IBM J. Res. Dev. **1**, 223 (1957); Philos. Mag. **21**, 863 (1970).
20. M. Büttiker and Y. Imry, J. Phys. C **18**, L467 (1985); M. Büttiker, Y. Imry, R. Landauer, and S. Pinhas, Phys. Rev. B **31**, 6207 (1985).
21. G. Cunibert, M. Sassetti and B. Kramer, Phys. Rev. B **57**, 1515 (1998).
22. J. Gabelli, J. Feve, J.-M. Berroir B. Placais, A. Cavanna, B. Etienne, Y. Jin and D. C. Glattli,Science **313**, 499 (2006).
23. J. Wang, B. G. Wang and H. Guo, Phys. Rev. B **75,** 155336 (2007).